\documentclass[twocolumn]{webofc}
\usepackage[varg]{txfonts}   
\usepackage{color}
\usepackage{amsmath}
\usepackage{bm}
\usepackage{dcolumn}
\def\Nabla{\bm{\nabla}}
\def\ve#1{{\bm{#1}}} 
\def\nuc#1#2#3{{}^{#2}_{#3}\mathrm{#1}}
\def\urm#1{\scriptstyle{\text{\textrm{\textmd{\textup{{#1}}}}}}}
\begin{document}
\title{Coulomb Energy Density Functionals for Nuclear Systems:
  Recent Studies of Coulomb Exchange and Correlation Functionals}
\author{
  \firstname{Tomoya} \lastname{Naito} \inst{1,2}\fnsep{\thanks{\email{naito@cms.phys.s.u-tokyo.ac.jp}}}
  \and
  \firstname{Ryosuke} \lastname{Akashi} \inst{1}
  \and
  \firstname{Gianluca} \lastname{Col\`{o}} \inst{3,4}
  \and
  \firstname{Haozhao} \lastname{Liang} \inst{2,1}
  \and
  \firstname{Xavier} \lastname{Roca-Maza} \inst{3,4}}
\institute{
  Department of Physics, Graduate School of Science, The University of Tokyo,
  Tokyo 113-0033, Japan
  \and
  RIKEN Nishina Center, Wako 351-0198, Japan
  \and
  Dipartimento di Fisica, Universit\`{a} degli Studi di Milano,
  Via Celoria 16, 20133 Milano, Italy
  \and
  INFN, Sezione di Milano,
  Via Celoria 16, 20133 Milano, Italy}
\abstract{%
  The Coulomb exchange and correlation energy density functionals for electron systems are applied to nuclear systems.
  It is found that
  the exchange functionals in the generalized gradient approximation
  provide agreements with the exact-Fock energy with one adjustable parameter within a few dozen $ \mathrm{keV} $ accuracy,
  whereas the correlation functionals are not directly applicable to nuclear systems due to the existence of the nuclear force.}
\maketitle
\section{Introduction}
\label{sec:intro}
\par
Atomic nuclei are composed of protons and neutrons that interact with one another through the nuclear and electromagnetic forces.
The former, which is much stronger than the latter, dominates the properties of atomic nuclei.
Nevertheless, in specific studies, it is crucial to evaluate contribution of the latter to the properties of atomic nuclei.
The mass difference of the mirror nuclei and energy of the isobaric analog state are such examples.
In this report, we focus on the recent studies of the Coulomb energy density functionals (EDFs)
of electron systems in the context of atomic nuclei 
\cite{Naito2018Phys.Rev.C97_044319,Naito2019Phys.Rev.C99_024309}
in the density functional theory (DFT) \cite{Hohenberg1964Phys.Rev.136_B864,Kohn1965Phys.Rev.140_A1133}.
\par
In the DFT for electron systems, the correlation energy is considered as well as the Hartree and exchange energies,
whereas in the nuclear DFT it is not considered explicitly.
The correlation EDF is tested for nuclear systems, where the local density approximation (LDA) is used,
and we have used the experimentally observed charge-density distribution for quantitative calculations of selected nuclei to avoid an error coming from the density \cite{Naito2018Phys.Rev.C97_044319}.
\par
To calculate Coulomb exchange energy, we carry out self-consistent Skyrme Hartree-Fock calculations by using the Perdew-Burke-Ernzerhof generalized gradient approximation (PBE-GGA) Coulomb exchange functional \cite{Perdew1996Phys.Rev.Lett.77_3865} instead of using the exact-Fock term,
and the optimal value of the free parameter $ \mu $ that appears in the PBE-GGA functional is also discussed \cite{Naito2019Phys.Rev.C99_024309}.
\section{Correlation Functional}
\label{sec:corr}
\par
The Coulomb correlation energies calculated by the charge density distribution \cite{DeVries1987At.DataNucl.DataTables36_495} in the LDA, $ E_{\urm{Cc}} $, for selected nuclei from light to heavy region are shown in Table \ref{tab:corr}.
For comparison, the Coulomb exchange energies calculated in the LDA, $ E_{\urm{Cx}} $, and the ratio $ E_{\urm{Cc}} / E_{\urm{Cx}} $ are also shown.
\par
It is seen that in these calculations, $ E_{\urm{Cc}} $ is all around $ 2 \, \% $ of $ E_{\urm{Cx}} $.
However, Bulgac and Shaginyan \cite{Bulgac1996Nucl.Phys.A601_103,Bulgac1999Phys.Lett.B469_1} evaluated that in atomic nuclei,
$ E_{\urm{Cc}}/E_{\urm{Cx}} $ would be around $ -40 \, \% $ to $ -20 \, \% $, instead of $ 2 \, \% $.
Hypothetically, if there is only Coulomb interaction
since correlation always further decreases the energy of the whole system,
we have the signs of the Hartree, exchange, and correlation energies as
$ E_{\urm{Cd}} > 0 $, $ E_{\urm{Cx}} < 0 $, and $ E_{\urm{Cc}} < 0 $, respectively.
In reality, the correlation EDF is not separable at all.
In Refs.~\cite{Bulgac1996Nucl.Phys.A601_103,Bulgac1999Phys.Lett.B469_1}
the correlation EDFs are written in terms of the response functions,
and such response functions are determined by the total interaction, i.e.,~mainly by the attractive nuclear part,
instead of the repulsive Coulomb part.
The total correlation energy is still negative,
mainly due to the contribution of the nuclear interaction.
As a result,
the contribution of Coulomb interaction becomes positive,
i.e.,~for the Coulomb energies, $ E_{\urm{Cc}} $ has the different sign as $ E_{\urm{Cx}} $.
In short, the correlation energy density functionals of electron systems cannot be applied directly to atomic nuclei.
It is also noted that the Coulomb correlation functional in the GGA gives around $ 30 $--$ 80 \, \% $ of $ E_{\urm{Cc}} $ in the LDA.
\begin{table}[!htb]
  \centering 
  \caption{
    Coulomb correlation energies $ E_{\urm{Cc}} $ for selected nuclei.
    Energies are shown in $ \mathrm{MeV} $.
    Data are taken from Ref.~\cite{Naito2018Phys.Rev.C97_044319}.}
  \label{tab:corr}
  \begin{tabular}{rD{.}{.}{4}D{.}{.}{5}D{.}{.}{6}}
    \hline \hline
    \multicolumn{1}{c}{Nuclei} & \multicolumn{1}{c}{LDA $ E_{\urm{Cx}} $} & \multicolumn{1}{c}{LDA $ E_{\urm{Cc}} $} & \multicolumn{1}{c}{$ E_{\urm{Cc}}^{\urm{LDA}} / E_{\urm{Cx}}^{\urm{LDA}} $} \\ \hline
    $ \nuc{O}{16}{} $ & -2.638 & -0.05218 & 1.978 \, \% \\
    $ \nuc{Ca}{40}{} $ & -7.087 & -0.1329 & 1.875 \, \% \\
    $ \nuc{Ca}{48}{} $ & -7.113 & -0.1332 & 1.873 \, \% \\
    $ \nuc{Ni}{58}{} $ & -10.28 & -0.1879 & 1.828 \, \% \\
    $ \nuc{Sn}{116}{} $ & -18.41 & -0.3361 & 1.826 \, \% \\
    $ \nuc{Sn}{124}{} $ & -18.24 & -0.3356 & 1.840 \, \% \\
    $ \nuc{Pb}{208}{} $ & -30.31 & -0.5524 & 1.823 \, \% \\
    \hline \hline
  \end{tabular}
\end{table}
\section{Exchange Functional}
\label{sec:exch}
\par
The GGA Coulomb exchange functionals have been proposed as 
\begin{equation}
  \label{eq:x-PBE}
  E_{\urm{Cx}}
  \left[
    \rho_{\urm{ch}}
  \right]
  =
  - \frac{3}{4}
  \frac{e^2}{4 \pi \epsilon_0}
  \left( \frac{3}{\pi} \right)^{1/3}
  \int
  \left[
    \rho_{\urm{ch}} \left( \ve{r} \right)
  \right]^{4/3}
  F \left(s \left( \ve{r} \right) \right)
  \, d \ve{r},
\end{equation}
where $ \rho_{\urm{ch}} $ is the charge density distribution, 
$ F $ is the enhancement factor depending on the density gradient,
and $ s $ denotes the dimensionless density gradient
\begin{equation}
  \label{eq:s}
  s = \frac{\left| \Nabla \rho_{\urm{ch}} \right|}{2 k_{\urm{F}} \rho_{\urm{ch}}},
  \qquad
  k_{\urm{F}}
  =
  \left(
    3 \pi^2 \rho_{\urm{ch}}
  \right)^{1/3}.
\end{equation}
In particular,
\begin{equation}
  \label{eq:F-PBE}
  F \left( s \right)
  =
  1 + \kappa
  -
  \frac{\kappa}{1 + \mu s^2 / \kappa},
\end{equation}
is used in the PBE-GGA Coulomb exchange functional
to satisfy some physical conditions,
and $ F \equiv 1 $ corresponds to the LDA one, 
i.e.,~the Hartree-Fock-Slater approximation \cite{Dirac1930Proc.Camb.Phil.Soc.26_376,Slater1951Phys.Rev.81_385}.
The parameter $ \kappa = 0.804 $ is determined for any value of $ \mu $ by the H\"{o}lder inequality.
In contrast, two different values of $ \mu $ have been widely used in the studies of atoms \cite{Perdew1996Phys.Rev.Lett.77_3865} and solids \cite{Perdew2008Phys.Rev.Lett.100_136406}, respectively.
For the PBE-GGA functional, $ \mu = 0.21951 $ is determined by the random phase approximation of the homogeneous electron gas.
Since this $ \mu $ can be a different value for nuclear systems, the free parameter of the PBE-GGA Coulomb exchange functional, $ \mu $, is multiplied by a factor $ \lambda $.
For the nuclear part, the SAMi functional \cite{Roca-Maza2012Phys.Rev.C86_031306} is used in the self-consistent calculation.
For comparison, the exact-Fock energies are also calculated \cite{Roca-Maza2016Phys.Rev.C94_044313}.
\par
\begin{figure}[!htb]
  \centering
  \includegraphics[width=1.0\linewidth]{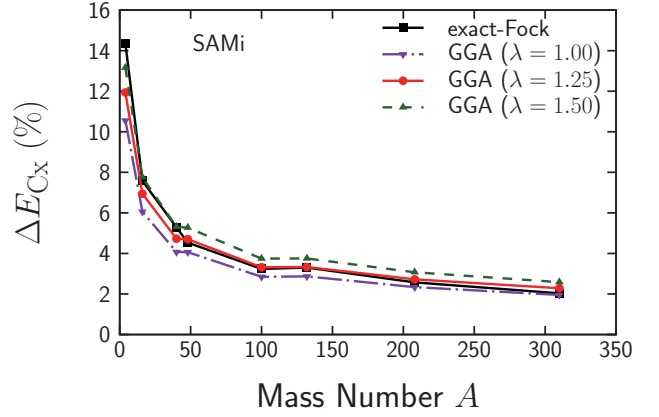}
  \caption{
    Deviation between the PBE-GGA and the LDA, $ \Delta E_{\urm{Cx}}^{\urm{LDA}} $
    defined as Eq.~\eqref{eq:ecx_diff}.
    Taken from Ref.~\cite{Naito2019Phys.Rev.C99_024309}.}
  \label{fig:systematic_dmagic}
\end{figure}

The deviation of the Coulomb exchange energy $ E_{\urm{Cx}} $ of PBE-GGA from that of LDA, $ \Delta E_{\urm{Cx}}^{\urm{LDA}} $,
\begin{equation}
  \label{eq:ecx_diff}
  \Delta E_{\urm{Cx}}^{\urm{LDA}}
  =
  \frac{E_{\urm{Cx}} - E_{\urm{Cx}}^{\urm{LDA}}}{E_{\urm{Cx}}}
\end{equation}
are shown as a function of mass number $ A $ in
Fig.~\ref{fig:systematic_dmagic}.
\par
It is found that in the light-mass region, to reproduce the exact-Fock results, $ \lambda = 1.50 $ or more is required,
while in the medium-heavy- and heavy-mass regions $ \lambda = 1.25 $ reproduces well the exact-Fock results.
The PBE-GGA result with $ \lambda = 1.00 $ reproduces the exact-Fock result in the case of the super-heavy nucleus $ \nuc{126}{310}{} $
since the ratio of the surface region to the volume region in the super-heavy nuclei is smaller than that in the medium-heavy or heavy nuclei.
\section{Conclusion}
\label{sec:conc}
\par
In these works \cite{Naito2018Phys.Rev.C97_044319,Naito2019Phys.Rev.C99_024309}, the Coulomb exchange and correlation EDFs in electron systems are applied to the nuclear systems.
On the one hand, the Coulomb correlation energy density functionals of electron systems are not applicable for atomic nuclei,
because these functionals are not separable and the nuclear interaction determines properties of atomic nuclei mainly.
On the other hand, the PBE-GGA Coulomb exchange functional with $ \lambda = 1.25 $ reproduces the exact-Fock energy in the self-consistent Skyrme Hartree-Fock calculations for atomic nuclei.
It should be emphasized that the numerical cost of the self-consistent calculations with the PBE-GGA exchange functional is $ O \left( N^3 \right) $, whereas that with the exact-Fock term is $ O \left( N^4 \right) $. 
\par
\begin{acknowledgement}
  TN and HL would like to thank the RIKEN iTHEMS program
  and the JSPS-NSFC Bilateral Program for Joint Research Project on Nuclear mass and life for unravelling mysteries of the $ r $-process.
  TN acknowledges the financial support from Computational Science Alliance, The University of Tokyo,
  Universit\`{a} degli Studi di Milano, 
  and the JSPS Grant-in-Aid for JSPS Fellows under Grant No.~19J20543.
  HL acknowledges the JSPS Grant-in-Aid for Early-Career Scientists under Grant No.~18K13549.
  GC and XRM acknowledge funding from the European Union's Horizon 2020 research and innovation program under Grant No.~654002.
\end{acknowledgement}
\bibliography{009_proc_NSD2019_Coulomb}
\end{document}